\documentclass[useAMS,usenatbib]{mn2e}
\usepackage{graphicx}
\usepackage{rotating}
\usepackage{wrapfig}
\usepackage{lscape}
\usepackage{longtable}
\usepackage{lscape}
\usepackage{amsmath}
\usepackage{overpic}
\usepackage[T1]{fontenc}
\usepackage{aecompl}
\usepackage[normalem]{ulem}
\usepackage[
	breaklinks,
        colorlinks=true,
        urlcolor=blue,		
        filecolor=blue,		
        citecolor=blue,		
        linkcolor=blue,		
        pdftitle={Untitled},
        pdfauthor={Your Name},
        pdfsubject={Just a test},
        pdfkeywords={test,t testing, testable stuff},
        pdfproducer={pdfLaTeX},
        pdfpagemode=None,
        hypertexnames=false,
        bookmarksopen=true]{hyperref}

\newcommand\nion[2]{#1\,\lowercase{{\sc #2}}}

\def\kmsec{${\rm km~s^ {-1}}$}
\def\BmV0{\mbox{$(B-V)^{\rm 0}$}}
\def\VmK0{\mbox{$(V-K)^{\rm 0}$}}
\def\MV0{\mbox{$M_{\rm V}^{\rm 0}$}}

\def\msun{${\rm M}_{\odot}$}

\def\simgt{\lower.5ex\hbox{$\; \buildrel > \over \sim \;$}}
\def\simlt{\lower.5ex\hbox{$\; \buildrel < \over \sim \;$}}
\def\teff{T_{\rm eff}}

\def\logg{\log~g}

\def\feh{\rm[Fe/H]}
\def\mgfe{\rm[Mg/Fe]}

\def\ezec{E_{\rm z}/E_{\rm c}}
\def\jzjc{J_{\rm z}/J_{\rm c}}

\def\EzJz{$[J_z/J_c,E_z/E_c]$}
\def\vpi{V_{\Pi}}
\def\vth{V_{\Theta}}
\def\vz{V_{\rm Z}}

\title[The Gaia-ESO Survey: a quiescent Milky Way]{The Gaia-ESO Survey: a quiescent Milky Way with no significant dark/stellar accreted disc\thanks{Based on data products from observations made with ESO Telescopes at the La Silla Paranal Observatory under programme ID 188.B-3002.  These data products have been processed by the Cambridge Astronomy Survey Unit (CASU) at the Institute of Astronomy, University of Cambridge, and by the FLAMES/UVES reduction team at INAF/Osservatorio Astrofisico di Arcetri. These data have been obtained from the Gaia-ESO Survey Data Archive, prepared and hosted by the Wide Field Astronomy Unit, Institute for Astronomy, University of Edinburgh, which is funded by the UK Science and Technology Facilities Council.}}

\author[Ruchti et al.]{G.~R.~Ruchti,$^{1}$\thanks{{\tt email: greg@astro.lu.se}} J.~I.~Read,$^{2}$ S.~Feltzing,$^{1}$ A.~M.~Serenelli,$^{3}$ P.~McMillan,$^{1}$ K.~Lind,$^{4}$
\newauthor
T.~Bensby,$^{1}$ M. Bergemann,$^{5}$ M.~Asplund,$^{6}$ A. Vallenari,$^{7}$ E.~Flaccomio,$^{8}$ 
\newauthor
E.~Pancino,$^{9,10}$ A.~J. Korn,$^{4}$ A.~Recio-Blanco,$^{11}$ A.~Bayo,$^{12}$ G.~Carraro,$^{13}$  
\newauthor
M.~T.~Costado,$^{14}$ F.~Damiani,$^{7}$ U.~Heiter,$^{4}$ A.~Hourihane,$^{15}$ P.~Jofr\'e,$^{15}$ G.~Kordopatis,$^{16}$   
\newauthor
C.~Lardo,$^{17}$ P.~de~Laverny,$^{11}$ L.~Monaco,$^{13,18}$ L.~Morbidelli,$^{19}$ L.~Sbordone,$^{20,21}$  
\newauthor
C.~C.~Worley,$^{15}$ S. Zaggia$^{7}$ \\
$^{1}$Lund Observatory, Department of Astronomy and Theoretical Physics, Box 43, SE-22100, Lund, Sweden\\
$^{2}$Department of Physics, University of Surrey, Guildford, GU2 7XH, Surrey, UK\\
$^{3}$Instituto de Ciencias del Espacio (ICE-CSIC/IEEC) Campus UAB, Carrer de Can Magrans, s/n x08193 Cerdanyola del Vallés\\
$^{4}$Department of Physics and Astronomy, Uppsala University, Box 516, SE-751 20 Uppsala, Sweden\\
$^{5}$Max-Planck Institut f\"{u}r Astronomie, K\"{o}nigstuhl 17, 69117 Heidelberg, Germany\\
$^{6}$Research School of Astronomy \& Astrophysics, Australian National University, Cotter Road, Weston Creek, ACT 2611, Australia\\
$^{7}$INAF - Padova Observatory, Vicolo dell'Osservatorio 5, 35122 Padova, Italy\\
$^{8}$INAF - Osservatorio Astronomico di Palermo, Piazza del Parlamento 1, 90134, Palermo, Italy\\
$^{9}$INAF - Osservatorio Astronomico di Bologna, via Ranzani 1, 40127, Bologna, Italy\\
$^{10}$ASI Science Data Center, Via del Politecnico SNC, 00133 Roma, Italy\\
$^{11}$Laboratoire Lagrange (UMR7293), Universit\'e de Nice Sophia Antipolis, CNRS,Observatoire de la C\^ote d'Azur,\\
CS 34229,F-06304 Nice cedex 4, France\\
$^{12}$Instituto de F\'isica y Astronomi\'ia, Universidad de Valparai\'iso, Chile\\
$^{13}$European Southern Observatory, Alonso de Cordova 3107 Vitacura, Santiago de Chile, Chile\\
$^{14}$Instituto de Astrof\'{i}sica de Andaluc\'{i}a-CSIC, Apdo. 3004, 18080 Granada, Spain\\
$^{15}$Institute of Astronomy, University of Cambridge, Madingley Road, Cambridge CB3 0HA, United Kingdom\\
$^{16}$Leibniz Institute for Astrophysics Potsdam, An der Sternwarte 16, 14482 Potsdam\\
$^{17}$Astrophysics Research Institute, Liverpool John Moores University, 146 Brownlow Hill, Liverpool L3 5RF, United Kingdom\\
$^{18}$Departamento de Ciencias F\'{i}sicas, Universidad Andr\'es Bello, Rep\'ublica 220, 837-0134 Santiago, Chile\\
$^{19}$INAF - Osservatorio Astrofisico di Arcetri, Largo E. Fermi 5, 50125, Florence, Italy\\
$^{20}$Millennium Institute of Astrophysics, Chile\\
$^{21}$Pontificia Universidad Cat\'{o}lica de Chile, Av. Vicu\~{n}a Mackenna 4860, 782-0436 Macul, Santiago, Chile
}

\date{Accepted 2015 April 8.  Received 2015 April 8; in original form 2015 March 11}

\begin{document}

\maketitle

\begin{abstract}
According to our current cosmological model, galaxies like the Milky Way are expected to experience many mergers over their lifetimes.  The most massive of the merging galaxies will be dragged towards the disc-plane, depositing stars and dark matter into an accreted disc structure.  In this work, we utilize the chemo-dynamical template developed in Ruchti et al. to hunt for accreted stars. We apply the template to a sample of 4,675 stars in the third internal data release from the Gaia-ESO Spectroscopic Survey.  We find a significant component of accreted halo stars, but find no evidence of an accreted disc component. This suggests that the Milky Way has had a rather quiescent merger history since its disc formed some 8-10 billion years ago and therefore possesses no significant dark matter disc.

\end{abstract}


\section{Introduction}\label{sec-intro}
In the standard $\Lambda$CDM cosmological model, galaxies build up their mass through the successive mergers of smaller galaxies. This process is known as hierarchical clustering \citep[e.g.][]{white1978,springel2006}.  It has long been a concern that the many mergers predicted by $\Lambda$CDM would destroy galactic discs \citep[e.g.][]{2008ApJ...683..597S}. Indeed, early simulations found excessive heating due to the bombardment of discs over cosmic time \cite[e.g.][]{eke2000}. More modern simulations have found, however, that the heating is dramatically reduced by moving to higher resolution \citep{kazantzidis2008,read2008} and if gas is included \citep{moster2010}. The very latest cosmological simulations now resolve the most massive sites of star formation ($\simlt  100$\,pc) where most of the stellar feedback energy is generated; these are beginning to produce more realistic discs, for the first time \citep[e.g.][]{2011ApJ...742...76G,2013MNRAS.436..625S,2014MNRAS.445..581H,agertz2014}. However, even at this resolution potentially ad-hoc fine-tuning is required to suppress over-cooling in supernovae shocks \citep[e.g.][]{2006MNRAS.373.1074S,2008MNRAS.387.1431D,2014MNRAS.445..581H,agertz2014,2015arXiv150105655K}. We are closing-in on the issue of whether $\Lambda$CDM can be reconciled with the large abundance of disc galaxies in the Universe today, but the question remains open \citep[e.g.][]{2015arXiv150207747S,2015arXiv150101311C}.


The Milky Way (MW) is a unique tool for understanding the formation and evolution of disc galaxies in the Universe. We can study its stellar populations -- which carry imprints of the history of the MW -- in greater detail than in any other galaxy. By measuring the chemistry, ages, and dynamics of its individual stars we can trace their origins. This is known as {\it Galactic Archaeology} \citep{eggen1962,freeman2002}. However, despite significant progress due to the advent of large stellar surveys like the Sloan Digital Sky Survey \citep[SDSS;][]{2011AJ....142...72E}, we still do not know the precise merger history of the Galaxy. This limits our ability to use the MW as a `rosetta stone' for galaxy formation, since we do not know if its merger history is typical or rare. 

\citet{minchev2014} argue that the relationship between stellar velocity dispersion and $\alpha$-abundance for stars from the RAdial Velocity Experiment \citep[RAVE,][]{kordopatis2013} suggest early massive mergers, with a declining merger intensity after disc formation some $\sim 9$ billion years ago; while \citet{deason2013} argue that an observed `break' in the density profile of the MW's stellar halo is also indicative of a quiet past, with few or no massive mergers over the past $\sim 9$\,billion years. Both of these studies rely on model interpretations of the data but are sensitive to mergers over the full Galactic history.   

Another approach is to look for moving groups. \citet{helmi2006} identified kinematic groups in the Galactic disc using the Geneva-Copenhagen Survey \citep{nordstrom2004}.  At least some of these are due to resonant orbits in the disc \citep[e.g. the Hercules stream;][]{bensby2007letter}.  However, some could have extragalactic origins \citep{stonkute2012,stonkute2013,zenoviene2014,zenoviene2015}.  The ages of the latter kinematic groups suggest they were accreted early on ($\simgt8$~Gyr ago). 

In this paper, we consider a new methodology that is most sensitive to massive $\sim 1:10$ mergers after disc formation, but is less reliant on models: hunting for accreted stars in the MW disc \citep[][hereafter R14]{ruchti2014}.  Furthermore, since we look for the {\it integral} of all debris in the disc, we have greater sensitivity than other methods. The idea is already presented in detail in R14; here we briefly summarise the key points. 

Merging satellite galaxies will deposit stars, gas, and dark matter into the MW \citep[e.g.][]{1996ApJ...460..121W,abadi2003,read2008}. For the most massive mergers, dynamical friction pulls the satellite towards the MW disc leading to an accreted `disc' of stars and dark matter \citep[e.g.][]{1989AJ.....98.1554L,read2008,2009MNRAS.397...44R,2009ApJ...703.2275P,2010JCAP...02..012L,pillepich2014}. These stars -- being born in shallower potential wells than the MW -- should also show distinct chemistry \citep{helmi2006,tolstoy2009,kirby2011,ruchti2014}.  

If we can find these `accreted disc stars', then they give us a powerful and direct probe of the merger history of the MW over the past $\sim 9$ billion years since the MW disc formed. The presence of such accreted disc stars encodes information about past massive mergers; its absence would suggest that such mergers did not occur. The accreted disc is also interesting because it contains dark matter -- the so-called `dark disc' \citep{1989AJ.....98.1554L,read2008,2009MNRAS.397...44R}. Such a dark disc, if present in the MW, has important implications for the local dark matter density and velocity distribution function. These in turn affect the expected dark matter signals in `direct' and `indirect' particle dark matter experiments (\citealt{bruch2009}; \citealt{2009PhLB..674..250B}; and for a review see \citealt{read2014}).

In R14, we developed a new method to hunt for accreted disc stars. We applied our methodology to a small sample of about $286$ stars with high-resolution abundances \citep{ruchti2011mpd,ruchti2013}, finding little evidence for any accreted disc stars in the solar neighbourhood.  We further demonstrated that this implies that the MW likely has a very light dark matter disc.  However, the sample we used was small and nearby ($<1$~kpc away), limiting the analysis to small-number statistics with strong observational and kinematical selection biases.  We thus concluded that a larger, unbiased sample was needed to confirm our results.

The Gaia-ESO Survey \citep{gilmore2012}, which began at the end of 2011, is an ongoing, five-year public spectroscopic survey conducted with the FLAMES/GIRAFFE and UVES instruments at the Very Large Telescope.  It will deliver high-quality spectra for $\sim100,000$ stars in the MW by 2017.  The recently released internal third data release (iDR3) contains nearly $20,000$ spectra for stars in the MW field as well as globular and open clusters.  Thus, the Gaia-ESO Survey presents an ideal data set with which to hunt for accreted stars in the MW.  In this paper, we utilise this extraordinary resource to expand our original analysis in R14.  

This paper is organised as follows. In Section~\ref{sec-background}, we briefly review the details of the chemo-dynamical template developed in R14 to identify accreted stars.  In Section~\ref{sec:obsdata}, we describe the spectroscopic data from the Gaia-ESO Survey, while the derivation of the physical and kinematic properties of the stars in our sample are detailed in Section~\ref{sec-analysis}.  In Section~\ref{sec:results}, we search for substructure in the Gaia-ESO sample, identifying accreted stars.   We then compute the mass and orbital parameters for the merging satellites in Section~\ref{sec-merge}.  Finally, in Sections~\ref{sec-qMW} and \ref{sec:conclusions}, we discuss the implications of our results and give concluding remarks, respectively.

\section{Background}\label{sec-background} 

In this section, we briefly summarise the main aspects of our chemo-dynamical template.  The reader is referred to R14 for more details.

\subsection{The kinematic template}

When satellite galaxies merge with the MW, they experience a dynamical friction force, given by:
\begin{equation} 
M_{\rm sat} \dot{\bf v} \simeq C \frac{\rho M_{\rm sat}^2}{v^3}{\bf v} 
\label{eqn:friction}
\end{equation} 
where $M_{\rm sat}$ is the mass of the infalling satellite galaxy; $\dot{\bf v}$ is the deceleration due to dynamical friction; $\rho$ is the background density (i.e. stars, gas, dark matter etc.); $C$ is some constant of proportionality; and $v = |{\bf v}|$ is the velocity of the satellite relative to the background \citep{1943ApJ....97..255C,2006astro.ph..6636R}.  From this equation, it can be determined that high-mass satellites will feel a larger force than low-mass satellites.  Further, a larger force is expected for encounters with the Galactic disc than the Galactic halo, since the former possesses a higher density of stars and gas.  Thus, higher-mass satellites that encounter the disc will be ``disc plane dragged'' towards the Galactic plane over time and will deposit stars into an accreted disc-like structure.


The accreted stars will be dynamically distinct from those stars that formed in the MW.  This was shown in R14 using simulations of a LMC-mass satellite galaxy undergoing a collisionless merger with a MW-mass host galaxy. We considered varying orbital eccentricity and inclination as in \citet{read2008}.  We investigated the location of the accreted and in situ stars in specific vertical angular momentum-energy space \EzJz, where for a Galactic potential $\Phi(R,Z)$ we have:
\begin{equation}
\frac{J_{\rm z}}{J_{\rm c}} = \frac{R \cdot V_{\Theta}}{R_{\rm c} \cdot V_{\rm c}}~{\rm \,\,\,;\,\,\,}~\frac{E_{\rm z}}{E_{\rm c}} = \frac{\frac{\vz^2}{2} + \Phi(R,Z)}{\Phi(R,0)},
\end{equation}
and $\vth$ and $V_{\rm Z}$ are the rotational and vertical velocity components with respect to the Galactic disc, respectively; $V_{\rm c}$ is the circular velocity, given by: 
\begin{equation}
V_{\rm c} = \sqrt{R_{\rm c} \cdot \left. \frac{\partial\Phi}{\partial R}\right|_{R=R_{\rm c}}};
\end{equation}
and $R_{\rm c}$ is the radius of a planar circular orbit with the same specific energy $E$ as the star. This is found by minimising:
\begin{equation}
\left| \frac{V_{\rm c}^2}{2} + \Phi(R_{\rm c},0) - E \right|.
\end{equation}
Note that a circular, in-plane orbit will have $\jzjc=\ezec=1$.

After the merger, the accreted population occupies regions of low-$\jzjc$ and low-$\ezec$, while the heated in situ stars primarily inhabit $\jzjc>0.8$ and $\ezec>0.95$. However, in R14 we found that the strongest kinematic differential between heated and accreted stars is really seen primarily in $\jzjc$. For this reason, in this paper we will focus from here on solely on the $\jzjc$ distribution.

\begin{figure*}
\vspace{-25mm}
\includegraphics[height=\textwidth,angle=-90]{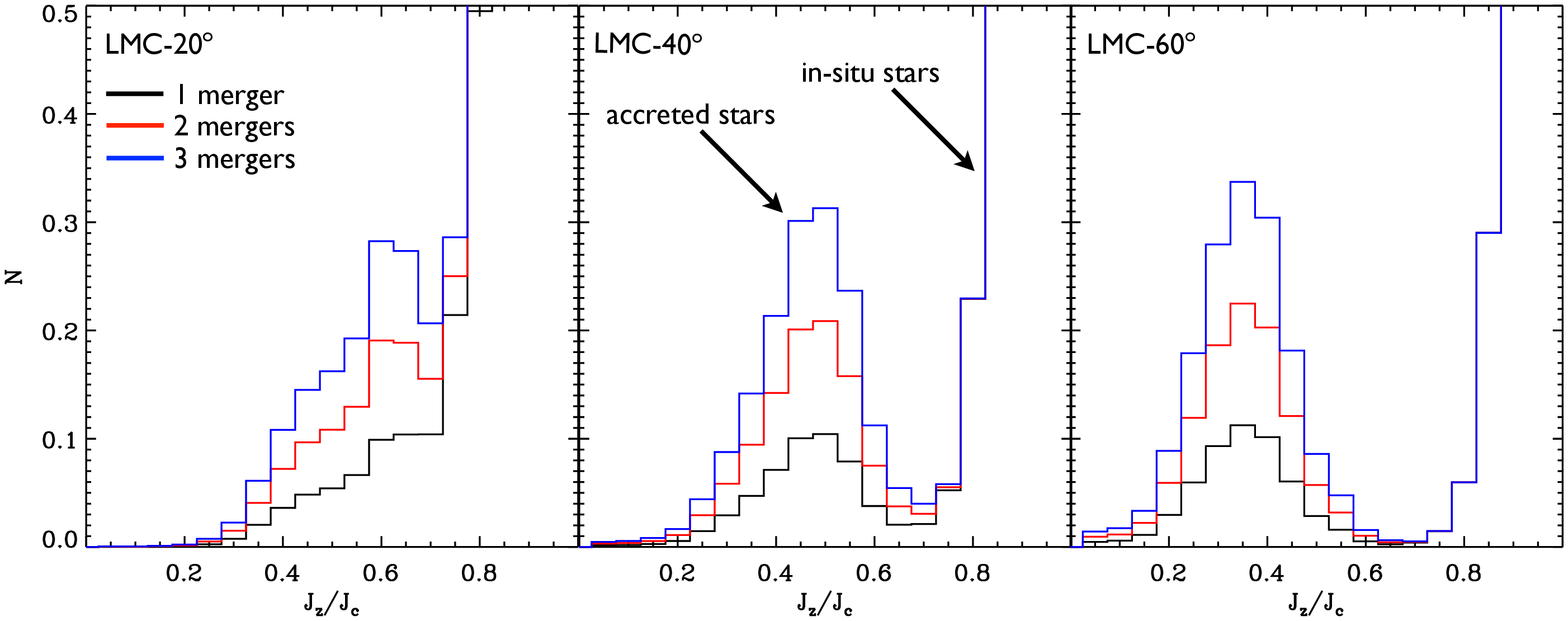}
\vspace{-28mm}
\caption{`Gaia-ESO volume' ($6 < R < 10$\,kpc; $|z| < 2$\,kpc) heated and accreted stars taken from a collisionless simulation of an LMC-like prograde merger with a MW-mass disc galaxy \citep{read2008}. The initial orbital inclination of the merging satellite was chosen to be 20$^\circ$ (left); 40$^\circ$ (middle) and 60$^\circ$ (right). The panels show histograms of $J_z/J_c$ (the ratio of the star specific angular momentum about the disc axis to that expected for a circular orbit at the guiding radius) for all stars assuming a single LMC-like merger (black) two identical such mergers where the stars are assumed to co-add linearly (red); and 3 such mergers (blue). The distributions are normalized such that their total area is equal to unity; they are all well-fit by a double Gaussian. In all cases, the accreted stars show up as a clear bump at lower $J_z/J_c$ than the in-situ star distribution. For very low inclination and/or low eccentricity mergers, however, the kinematics of the accreted and in-situ stars can overlap. For this reason, chemistry becomes important since the metallicity and abundance of stars born in a satellite should differ from those born in-situ in the Milky Way disc (see \S\ref{sec:chemtemplate}).}
\label{fig-accsim}
\vspace{0mm}
\end{figure*}

In Figure \ref{fig-accsim}, we show $\jzjc$ distributions from a simulation where a single LMC-like satellite merged on a prograde orbit inclined at 20$^\circ$ (left); 40$^\circ$ (middle) and 60$^\circ$ (right) to a Milky-Way like galaxy disc. The simulation was as outlined in R14 and taken from the simulation suite already published in \citet{read2008}. In practice, we may expect several such mergers, each on a different orbit, leading to a broader or multi-peaked histogram rather than one clean near-Gaussian peak as in Figure \ref{fig-accsim}. Here, we crudely estimate the effect of these multiple mergers by assuming identical mergers that linearly co-add stars. These are shown by the black (single); red (2-mergers) and blue (3-mergers) lines.

The behaviour for LMC-mass mergers seen in Figure \ref{fig-accsim} is distinct from lower mass mergers. As we move to lower mass satellites, we expect many more accretion events with a nearly isotropic distribution \citep{read2008}. Disc plane dragging by dynamical friction becomes less effective as $M_{\rm sat}$ shrinks (see equation \ref{eqn:friction}) and thus stars accreted from these low-mass satellite galaxies will have on average very low specific angular momenta, peaking around $\jzjc\sim0.0$. Such stars will contribute primarily to the canonical `stellar halo', rather than the `accreted disc'.

Based on these dynamical models, we concluded in R14 that `accreted disc stars' should occupy regions where $0.2<\jzjc<0.8$, while stars accreted from small satellites typically end up as accreted halo stars (and/or streams) with $\jzjc<0.2$. 

It is clear from Figure \ref{fig-accsim} that even a single LMC merger at $20^\circ$ inclination shows up as a clear `bump' in the $\jzjc$ histogram. However, for even lower inclination angle, the merger debris will become increasingly buried in the tail of the Milky Way in-situ stars. As discussed in R14, the solution to this is to study also the distinct {\it chemistry} of accreted stars.

\subsection{The chemical template}\label{sec:chemtemplate}

Stars born in small potential wells form from gas typically at lower surface density. This leads to a lower star formation efficiency and therefore lower alpha-to-iron ratios for a given metallicity bin (e.g. $\mgfe$\footnote{The bracket notation is defined as $[{\rm X}/{\rm Y}]\equiv\log(n({\rm X})/n({\rm Y}))_{\star}-\log(n({\rm X})/n({\rm Y}))_{\odot}$ for any elements X and Y.}). This can be understood from simple chemical evolution models, but is also empirically observed in the surviving MW satellites \citep[][R14]{matteucci1990,tolstoy2009}. It has also been shown that higher-mass satellites reach higher average metallicities (and possibly higher alpha-to-iron ratios for the largest masses) than lower-mass satellites \citep[e.g.][]{lanfranchi2006,kirby2013}.  In R14, it was shown that stars in present-day surviving dwarf galaxies, as well as the LMC, typically have $\mgfe\simlt0.3$ above $\feh\simgt-1.5$ (see their Fig.~1). We therefore expect that approximately LMC-mass mergers that will be significantly disc plane dragged, will also have $\mgfe \simlt 0.3$ at metallicities above $\feh\sim-1.5$.

The combination of chemistry and kinematics is required to unambiguously detect accreted stars in the disc. Stars can also be born in-situ at low efficiency and low metallicity in the outer disc, migrating later on into the solar neighbourhood. Similarly, accreted stars from satellites on particularly planar, circular, orbits could masquerade kinematically as in-situ stars (R14). For these reasons, it is the combination of `hot' stellar kinematics and low alpha abundance that provides the smoking-gun evidence for accreted stars that we searched for in R14 and that we will search for in this paper.


In R14, we applied our analysis to a small sample of 286 stars, and restricted our search for accreted stars to only regions where $\mgfe<0.3$.  However, with a large enough sample, we can search for sub-structure in the distribution of $\jzjc$ over all the metallicity-abundance space.  In the present work, we use a sample from the Gaia-ESO Survey that is more than ten times the size of the sample in R14 to search for accreted stellar components in the MW. 

\section{Observational Data}\label{sec:obsdata}

In this paper, we make use of the iDR3 data from the Gaia-ESO Survey for $7,723$ MW field stars with the GIRAFFE spectrograph, which have been observed with both the HR10 ($R\sim19,800$) and HR21 ($R\sim16,200$) settings.  All targets were selected according to their colour and magnitude, using the Visible and Infrared Survey Telescope for Astronomy (VISTA) photometry.  The selection function of the Gaia-ESO Survey will be presented in a future paper (Gilmore et al., in prep.), while some details can be found in \citet{recioblanco2014}.  


Radial velocities are measured from the GIRAFFE spectra, with typical uncertainties of $\sim0.3$~\kmsec~(see Koposov et al. in prep., for more details).  The stellar effective temperature ($\teff$), surface gravity ($\logg$), and metallicity ($\feh$) were derived by the Gaia-ESO Survey working group in charge of the GIRAFFE spectrum analysis for FGK stars (WG10).  The details of this analysis will be presented in Recio-Blanco et al. (in prep.).  Briefly, the spectra were analysed by three research groups using different spectral analysis codes: MATISSE \citep{recioblanco2006}, FERRE \citep{allendeprieto2006}, and SME \citep{valenti1996}.  

All analyses were performed using the 1D LTE spherically-symmetric ($\logg<3.5$) and plane-parallel ($\logg\geq3.5$) MARCS model atmospheres \citep{gustafsson2008} and a common atomic (and molecular) line list (Heiter et al. in prep.).  The results from the three different groups were corrected for method-to-method offsets using multi-linear transformations as a function of $\teff$ and $\logg$, and then averaged to give a final estimate of the stellar parameters for each spectrum.   The stellar parameters were also calibrated with respect to several benchmark stars \citep{jofre2014}.  Subsequently, individual abundances were estimated from the spectra and averaged amongst the three methods.  

For our analysis, we make particular use of [Fe/H] and the [Mg/Fe] ratio.  We thus utilized the measured Fe and Mg abundances from \nion{Fe}{I} and \nion{Mg}{I} lines in the GIRAFFE spectra, respectively.  Moreover, we averaged the results for several measurements of the solar spectrum to compute the solar abundances.  We found average values of $A({\rm Fe})_{\odot}=7.51$ and $A({\rm Mg})_{\odot}=7.60$.  These values are very similar to the solar abundances measured by \citet{asplund2009}, which are the solar abundances used in R14.  We therefore used these solar values when computing $\feh$ and $\mgfe$ in order to place the data on similar scales to R14.

Since the quality of the final stellar parameters depends on the signal-to-noise ratio (SNR) of the spectrum, we selected only stars with ${\rm SNR}\geq15$ in both HR10 and HR21.  Additionally, to reduce possible abundance scattering in our analysis, we further restricted our sample to only those stars with $\mgfe$ uncertainties of  $\delta_{\mgfe}<0.2$.  We further removed four stars that did not have a measured Mg abundance.  Finally, we removed any stars that do not have 2MASS or VISTA magnitude estimates and UCAC4 \citep{zacharias2013} or PPMXL \citep{roeser2010} proper motions, since these are important for deriving the kinematics of the stars (see \S\ref{sec-kin}).

\begin{figure}
\centering
\includegraphics[height=0.45\textwidth]{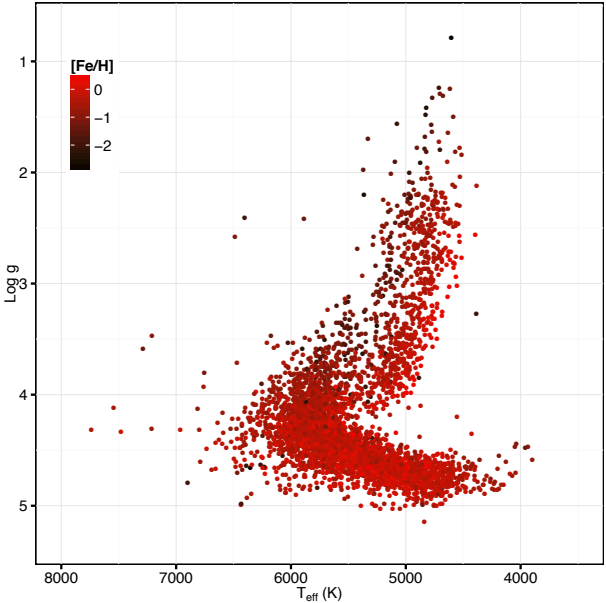}
\caption{Location of our sample, consisting of $4,675$ stars, in the Hertzsprung-Russell diagram.  The stars are color-coded according to their measured $\feh$.}
\label{fig-hrd}
\end{figure}

Our final sample consists of $4,675$ stars, with average uncertainties in their final stellar parameter estimates of $\left<\delta_{\teff}\right>\sim68\pm75$~K, $\left<\delta_{\logg}\right>\sim0.13\pm0.11$, and $\left<\delta_{\feh}\right>\sim0.06\pm0.03$, while $\delta_{\mgfe}\sim0.05\pm0.02$.  These uncertainties were estimated from the method-to-method scatter in the combination process.  Figure~\ref{fig-hrd} shows the Hertzsprung-Russell (H-R) diagram for our final data sample.

\section{The Dynamical Analysis}\label{sec-analysis}

\subsection{VISTA vs. 2MASS magnitudes}\label{sec-mags}

For most stars, both VISTA and 2MASS magnitudes were available in the Gaia-ESO internal data archive.  The VISTA photometry typically has a higher precision than 2MASS, especially for dim stars (mean uncertainty in $K_{\rm S}$ is $\sim0.03$~mag versus $\sim0.06$~mag, respectively).  Thus, we adopted the VISTA  $J_{\rm v}$ and $K_{\rm S,v}$ magnitudes when available for a given star.  There were four cases in which no VISTA magnitude was listed.  For these, we adopted the 2MASS magnitude.  

Our derivation of the distance to the stars requires a 2MASS $K_{\rm S,2}$ magnitude (see \S\ref{sec-mage}).  Thus, for stars in which we adopt the VISTA photometry, we converted the VISTA $J_{\rm v}$ and $K_{\rm S,v}$ magnitudes to the 2MASS $J_{\rm 2}$ and $K_{\rm S,2}$ magnitudes using:
\begin{eqnarray}
K_{\rm S,2} & = & K_{\rm S,v} - 0.01081~(J_{\rm v}-K_{\rm S,v}) \\
J_{\rm 2} & = & J_{\rm v} - 0.07027~(J_{\rm v}-K_{\rm S,v}).
\end{eqnarray}
These equations were derived from those that were determined to account for zero-point offsets between the VISTA and 2MASS magnitudes\footnote{For more details, see: http://casu.ast.cam.ac.uk/surveys-projects/vista/technical/photometric-properties}.

Although we adopt a combination of VISTA and 2MASS photometry in order to maximise the size of our sample, our overall results do not change if we adopt one or the other, exclusively.

\subsection{Absolute magnitudes and ages}\label{sec-mage}

Both the age and absolute magnitudes $M_{K_{\rm S,2}}$ and $M_{J_{\rm 2}}$ of each star in our sample were inferred from the estimated stellar parameters ($\teff$, $\logg$, and $\feh$) and a library of evolutionary tracks by using BeSPP (Bellaterra Stellar Parameter Pipeline). A detailed description of the Bayesian method implemented in BeSPP can be found in \citet{serenelli2013}. 

In this work, we use a newly computed grid of stellar models using Garstec \citep{weiss2008}. While most of the physical input in the models is the same as  described in \citet{serenelli2013} there is one important exception. For the new grid of models, certain amount of extra-mixing below the convective envelopes has been added following a phenomenological recipe similar to that given by \citet{vandenberg2012}. Further details will be discussed elsewhere (Serenelli 2015, in prep.). 

A solar model has been calibrated to fix the mixing length parameter for all the grid models ($\alpha_{\rm MLT}= 1.798$). The initial composition of the solar model is $Z_\odot=0.01940$ and $Y_\odot=0.2712$. 
For the whole grid we choose the GN93 solar composition \citep{grevesse1993} because it leads to solar interior models that are in agreement with helioseismic data.  As shown by \citet{jcd2009}, this choice leads to a correct estimation of the solar age, in contrast to models using the more recent AGSS09 solar composition \citep{asplund2009} which overestimate the solar age by about 10\%. 

The grid of stellar models spans the mass range $0.6\leq M/M_{\odot}\leq3.0$ with steps of $\Delta M=0.01~M_{\odot}$ and the metallicity range $-3.0\leq\feh\leq0.6$ with steps of $\Delta\feh=0.1$ for $\feh\leq0$ and $\Delta\feh=0.05$ for $\feh>0$. The initial composition of models is fixed by assuming a constant $\Delta Y/\Delta Z= 1.17$ as determined by anchoring this relation to the solar model and the helium abundance $Y_{\rm SBBN}= 0.2485$ determined from standard Big Bang Nucleosynthesis \citep{steigman2007}. 

As it has been originally discussed by \citet{salaris1993}, $\alpha$-enhancement can be mimicked in solar-scaled stellar tracks by modifying $\feh$ by an amount proportional to the $\alpha$-enhancement.  Based on a subset of stellar tracks including $\alpha$-enhancement, we find the following relation for the adjusted $\feh$
\begin{equation}
\feh_{\rm adjusted} = \feh + 0.22~\mgfe / 0.4,
\end{equation}
where $\mgfe$ is used as a proxy for the $\alpha$-enhancement of the star. 
Therefore, in BeSPP we effectively account for $\alpha$-enhancement by fitting $\teff$, $\logg$, and $\feh_{\rm adjusted}$.

In \citet{serenelli2013}, we showed that the stellar mass and age cannot be determined for stars with $\logg<3.5$.  This is due to the degeneracy of the stellar evolutionary tracks in the $\teff-\logg$ plane at these surface gravities.  Therefore, in this work we only present ages for stars with $\logg>3.5$.

From these fits, we obtain a full probability distribution function (PDF) for the absolute $M_{K_{\rm S,2}}$ magnitude and age of a given star.  These PDFs are not Gaussian for all stars.  We can thus use these PDFs to determine subsequent PDFs for the kinematic and orbital information for the stars.

\subsection{Distances, kinematics, and orbital parameters}\label{sec-kin}

The distances, kinematics, and orbital parameters of the stars were derived using a Monte Carlo approach.  First, we constructed PDFs for the $J_{\rm 2}$ and $K_{\rm S,2}$ magnitudes (described in \S\ref{sec-mags}), proper motions, and radial velocities (from the Gaia-ESO Survey) assuming Gaussian distributions with sigma equal to the uncertainty in the given quantity.  For a single star in our sample, we next created 3000 Monte Carlo sets of these parameters drawn from these PDFs, as well as the (possibly non-Gaussian) PDFs for the absolute $M_{K_{\rm S,2}}$ magnitude.  We then derived the distance, the 3D space motions, and orbital parameters for each Monte Carlo parameter set.  Each of these derivations thus has an associated PDF.  Below we give further details for each of the parameters.

\subsubsection{Distance}

The distance was determined following the iterative procedure described in \citet{ruchti2011mpd}.  This method uses $K_{\rm S,2}$ and $M_{K_{\rm S,2}}$ to derived an initial distance estimate using the distance modulus.  The reddening is then estimated from the \citet{schlegel1998} dust maps.  Values of $E(B-V)>0.1$ were corrected according to the formula given in \citet{bonifacio2000}: $E(B-V)_{\rm corrected}=0.035+0.65*E(B-V)_{\rm Schlegel}$.  Finally, we assume that for distances within the dust layer (assumed to have a scale-height of $h=125$~pc), the reddening at a distance $D$ and Galactic longitude $b$ will be reduced by a factor $1-\exp(-|D\sin b|/h)$. This reduction only affects those stars that are relatively nearby and lie close to the Galactic plane.  The $E(B-V)$ reddening was then converted to the $K_{\rm S,2}$ extinction coefficient, given by $A_{K_{\rm S,2}}=0.366*E(B-V)$ \citep{casagrande2014}.  The mean value of $A_{K_{\rm S,2}}$ was less than $0.1$~mag, with an average uncertainty of $\sim0.1$~mag.  This corresponds to a $\simlt5\%$ uncertainty in the distance.  The distance estimate and reddening were iterated until there was a difference in distance of $1\%$ between two consecutive iterations.

\subsubsection{Kinematics and orbital parameters}

We next combined the distance with the proper motions and radial velocities to derive the full 3D space motions in Galactocentric cylindrical coordinates ($\vpi$, $\vth$, $\vz$).  We adopted UCAC4 proper motions for the stars.  However, when unavailable (this is especially true for the more distant stars), we used the PPMXL proper motions.  Although there are possible systematics between the two sets of proper motions \citep[see][]{zacharias2013}, the overall results of our analysis do not change if we adopt one or the other.   

The average uncertainties in the proper motions are $\sim5~{\rm mas~yr^{-1}}$ and $\sim7~{\rm mas~yr^{-1}}$ for UCAC4 and PPMXL, respectively.  The velocities were corrected to the solar velocity with respect to the local standard of rest (LSR), given by $(U_{\odot}, V_{\odot}, W_{\odot})=(14.00, 12.24, 7.25)$~\kmsec{} from \citet{schonrich2010} and \citet{schonrich2012}.  An additional $V_{\rm LSR}=220$~\kmsec{} was added to put the space motions in the Galactic rest frame.

We next computed the orbital parameters of the stars following the methodology described in R14, which assumes a three component Galactic potential: a \citet{hernquist1990} bulge, a \citet{miyamoto1975} disc, and a \citet{navarro1996} dark matter halo.  For these calculations, we assumed the Sun sits at $R_{\odot}=8$~kpc.  We computed the specific angular momentum ($\jzjc$), as defined in \S\ref{sec-background}.

The advantage of using $\jzjc$ instead of other orbital parameters, such as the orbital eccentricity, is that it is less dependent on the chosen Galactic potential.  For example, we have adopted the masses and constants for each potential from \citet{gomez2010}.  However, it is possible that the bulge mass was over-estimated \citep[e.g.][]{laurikainen2010}.  If we lower the mass of the bulge by half, this leads to differences of less than 0.05 in $\jzjc$.

\subsubsection{Final PDFs and uncertainty estimates}\label{sec-pdfs}

Following the procedures described in the previous sections, we then combined the estimates for all 3000 Monte Carlo parameters sets to construct the PDFs for the distance, space motions, and $\jzjc$ for each star in our sample.    
The best estimate for each quantity was taken to be that at which the PDF is maximised.  The lower and upper uncertainties were adopted to be the values a the boundary of the $68.3\%$ confidence interval surrounding the value at the maximum.  

This resulted in typical upper and low age uncertainties of $\sim4$~Gyr and $\sim1$~Gyr, respectively.  The typical uncertainty in the distance was found to be $\sim10$\%.  This, combined with uncertainties in the proper motions and radial velocities, led to an the uncertainty in the space velocities that is typically $<50$\kmsec.  All of this together leads to mean upper and lower uncertainties in $\jzjc$ of $\sim0.11$ and $\sim0.08$, respectively. 

Larger uncertainties in the distance and proper motions often lead to increased uncertainty in $\jzjc$.  For many stars, however, the PDFs for the distance are not necessarily Gaussian.  This is also the case for the age PDFs.  Further, correlations among the different uncertainties are very non-linear and complex.  For example, the distance and proper motions are determined independently.  Thus, larger uncertainties in the distance do not necessarily correspond to larger uncertainties in the proper motions.  It is therefore important to retain the information in the derived PDFs for $\jzjc$ and ages during our analysis.  

\subsection{Aside: the Gaia-ESO selection function}

In R14, our sample was quite small and kinematically biased due to the selection function in the \citet{ruchti2011mpd} study.  Our current sample, however, is more than ten times the size of the R14 sample.  Moreover, there is no kinematic selection for the MW fields in the Gaia-ESO Survey.  Instead, the stars are selected according to their colour and magnitude.   This can cause some bias in the age$-$metallicity relation of the stars, in the sense that young, metal-poor stars and old, metal-rich stars are under-sampled \citep[e.g.][]{bergemann2014}.  Further, the selection means that the stars in our sample probe different volumes, with dwarfs and giants probing smaller and larger volumes, respectively.  This means that our sample is not volume complete, and so our sample may over- (or under-) populate stars from different regions of the MW.  This can cause a bias in the overall kinematic distributions and number density of stars in relation to the distance from the Sun.  This is due to the fact that the typical motions of the stars in the MW differ, depending on their position in the Galaxy.

Given our sample is not volume-complete, we cannot easily quantify the relative number density of accreted stars versus {\it in situ} stars within a specific volume.  However, as shown in R14, stars identified as accreted stars in our chemo-dynamical analysis are all genuine accreted stars, even for a more complicated selection function, which includes kinematic biases.  Since there is no kinematic selection in the Gaia-ESO Survey, it will have a small effect on our present analysis.  We therefore can study the dynamical distributions to identify accreted stars in our sample without any underlying corrections to the models presented in \S\ref{sec-background}.

\section{Hunting for substructure}\label{sec:results}    

\subsection{Abundance trends of the low $\jzjc$ stars}\label{sec-abund}

The size of our present data set from the Gaia-ESO Survey allows us to search for dynamical substructure throughout the $\mgfe-\feh$ space.  As an initial exploration, we first looked at the abundance trends of the stars that have orbits consistent with possible accretion, i.e. $\jzjc<0.5$ (see Figure~\ref{fig-accsim}).   

\begin{figure}
\centering
\includegraphics[height=0.6\textwidth]{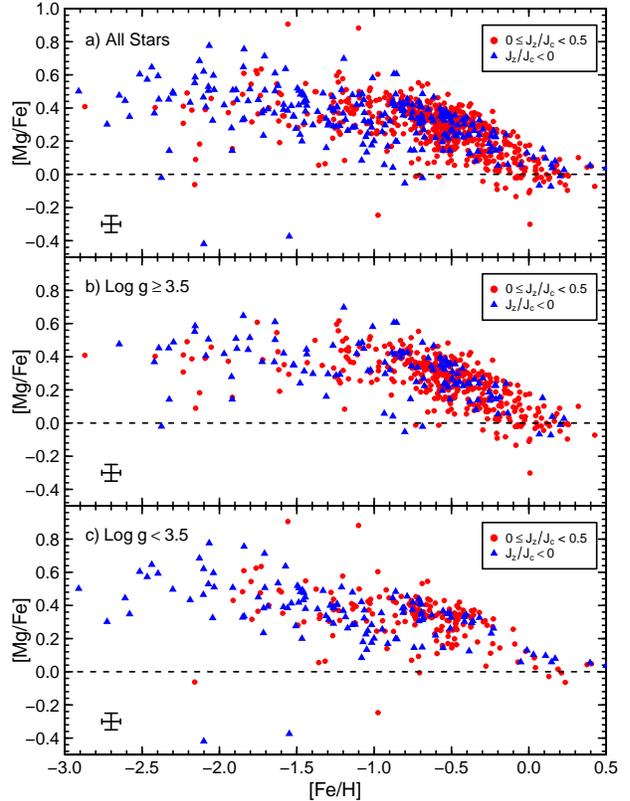}
\caption{Location of stars in the $\mgfe-\feh$ plane, highlighting stars with $\jzjc<0$ (blue triangles) and those with $0\leq\jzjc<0.5$ (red circles).  (a) All stars in our sample.  (b) Only stars with $\logg\geq3.5$, defined as sub-giants/dwarfs.  (c) Only stars with $\logg<3.5$, defined as giants.  Stars with $\jzjc>0.5$ are not plotted. Typical uncertainties are shown in the lower left-hand corners.}
\label{fig-mgsplit}
\end{figure}

In Figure~\ref{fig-mgsplit}, we plot our sample in the $\mgfe-\feh$ plane, highlighting those stars with $\jzjc<0$ (blue) and $0\leq\jzjc<0.5$ (red).  We further separate the sample into dwarfs/sub-giants ($\logg\geq3.5$) and giants ($\logg<3.5$), which are shown in Figure~\ref{fig-mgsplit}b and \ref{fig-mgsplit}c, respectively.

\subsubsection{Retrograde stars ($\jzjc < 0$)}

The stars with $\jzjc<0$ (i.e. those moving retrograde with respect to the disc) are $\alpha$-enhanced ($\left<\mgfe\right>\sim0.4-0.5$) up to a metallicity of $\sim-1.5$~dex. From there, their average [Mg/Fe] ratios predominantly decrease with increasing metallicity, with stars reaching $\mgfe\simlt0.2$ at $\feh\sim-0.8$. This is most clearly seen in Figures~\ref{fig-mgsplit}a and \ref{fig-mgsplit}c. (There are also some stars with $0<\jzjc<0.5$ that also follow this trend, albeit with a much larger spread.)

\begin{figure*}
\centering
\includegraphics[height=0.33\textwidth]{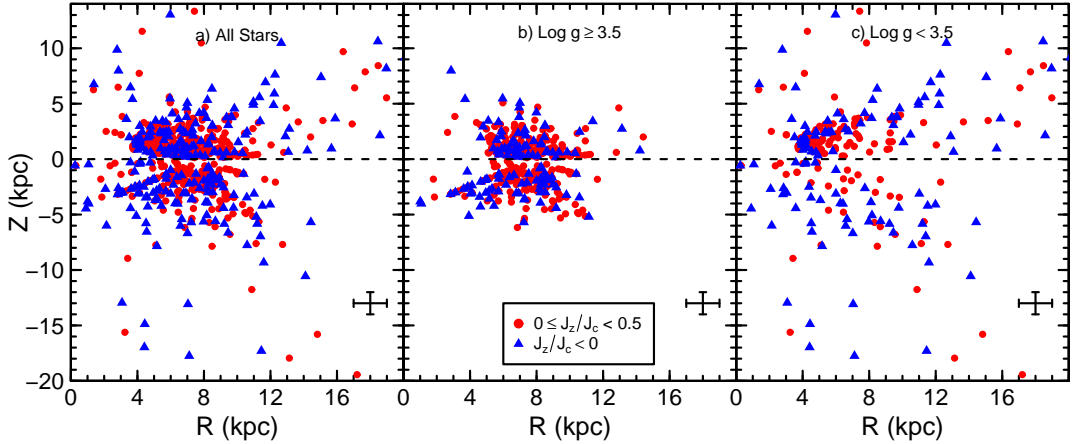}
\caption{Position of stars in the Galactocentric $R-Z$ plane, highlighting stars with $\jzjc<0$ as blue, triangles and those with $0\leq\jzjc<0.5$ as red, circles.  (a) All stars in our sample.  (b) Only stars with $\logg\geq3.5$, defined as sub-giants/dwarfs.  (c) Only stars with $\logg<3.5$, defined as giants.  Typical uncertainties are less than $1$~kpc in both $R$ and $Z$.  For the most distant stars, the uncertainties can range up to $\simgt2$~kpc.}
\label{fig-rzpos}
\end{figure*}

The metallicity behaviour of our $\jzjc < 0$ stars is qualitatively similar to the trends found for surviving dwarf galaxies \citep[e.g.][]{tolstoy2009}.  These stars also have abundances and kinematics that fit within our definition of accreted stars in R14 and \S\ref{sec-background}, suggesting an accretion origin. Moreover, these trends are very similar to those found by \citet{nissen2010}, in which they investigate high-velocity stars in the solar neighbourhood. Finally, note that the majority of the stars in our sample with $\jzjc<0$ sit at distances {\it outside} of the solar neighbourhood ($Z>1$~kpc or $R<7$~kpc or $R>9$~kpc).  This is shown in Figure~\ref{fig-rzpos}, where the positions of the stars are plotted in the Galactocentric $R-Z$ plane.  This further strengthens the conclusion that these are stars accreted from low mass dwarf galaxies. It also proves that the results found by \citet{nissen2010} are not exclusive to the solar neighbourhood.

\subsubsection{Low angular momentum stars ($\jzjc < 0.5$)}

A second interesting trend can be identified at metallicities between $-0.8$ and $-0.3$~dex.  Here, there appears to be a possible over-density of stars with $\jzjc<0.5$ (both red and blue symbols in Figure~\ref{fig-mgsplit}) and $\mgfe\sim0.3-0.4$. This is best seen in the giants (Figure~\ref{fig-mgsplit}c), showing also a slight decrease with increasing metallicity. This clump will be discussed further in \S\ref{sec-acc1}.

\subsection{Identifying substructure in $\jzjc$}\label{sec-acc1}

In the previous section, we found stars that have both $\mgfe$ ratios and specific angular momenta compatible with our chemo-dynamical analysis in \S\ref{sec-background}. In addition, there may be an over-density of low-$\jzjc$ stars at higher metallicity and high-$\alpha$.  However, we have not yet determined the statistical significance of these features. Since they lie in the tail of the thin/thick disc of the Milky Way, they could be contaminated by thin/thick disc stars, or even completely spurious.

\begin{figure}
\centering
\includegraphics[height=0.33\textwidth]{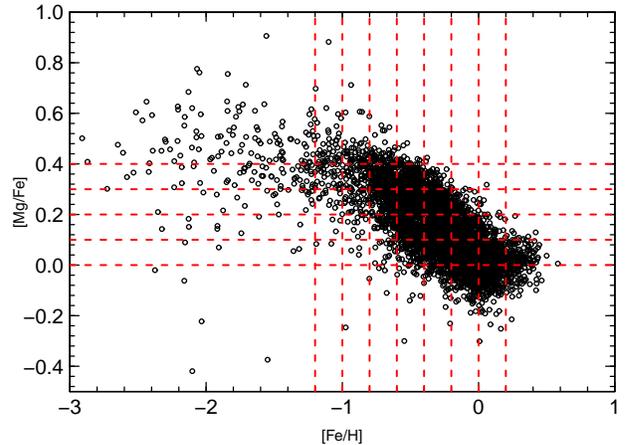}
\caption{Location of our stars in the $\mgfe-\feh$ plane.  The grid of bins used in the first step in the analysis in \S\ref{sec:results} is shown by the red, dashed lines.  Note, at the extrema of the grid, the bins extend to infinity.}
\label{fig-mgfe}
\end{figure}

To determine if these stars can be attributed to significant accreted components, and to account for uncertainties in their $\jzjc$ values, in this section we investigate their detailed distributions in $\jzjc$\footnote{In addition to $\jzjc$, R14 also computed the vertical specific energy ($\ezec$) for each star. In that work, we found that no distinguishing information could be extracted from the distribution in $\ezec$ (see, e.g., Fig.~7 of R14). Thus, for the present analysis, we concentrate on $\jzjc$ only.}. This was initially done by slicing our data into a grid of 0.2~dex and 0.1~dex bins in $\feh$ and $\mgfe$, respectively.  These were chosen so that the bin size in $\mgfe$ was about twice the size of the uncertainty, while the bin size in $\feh$ was chosen to retain significant number statistics ($n_{\rm stars}>10$) in the majority of the metal-poor bins.  Shifting the bins by $0.05$~dex in any direction has no major impact on subsequent results.  The grid is shown in Figure~\ref{fig-mgfe}.  At the extremes of both ranges, all stars are included to ``infinity'' in order to improve the statistics.  

\begin{figure*}
\centering
\includegraphics[angle=0,height=0.7\textwidth]{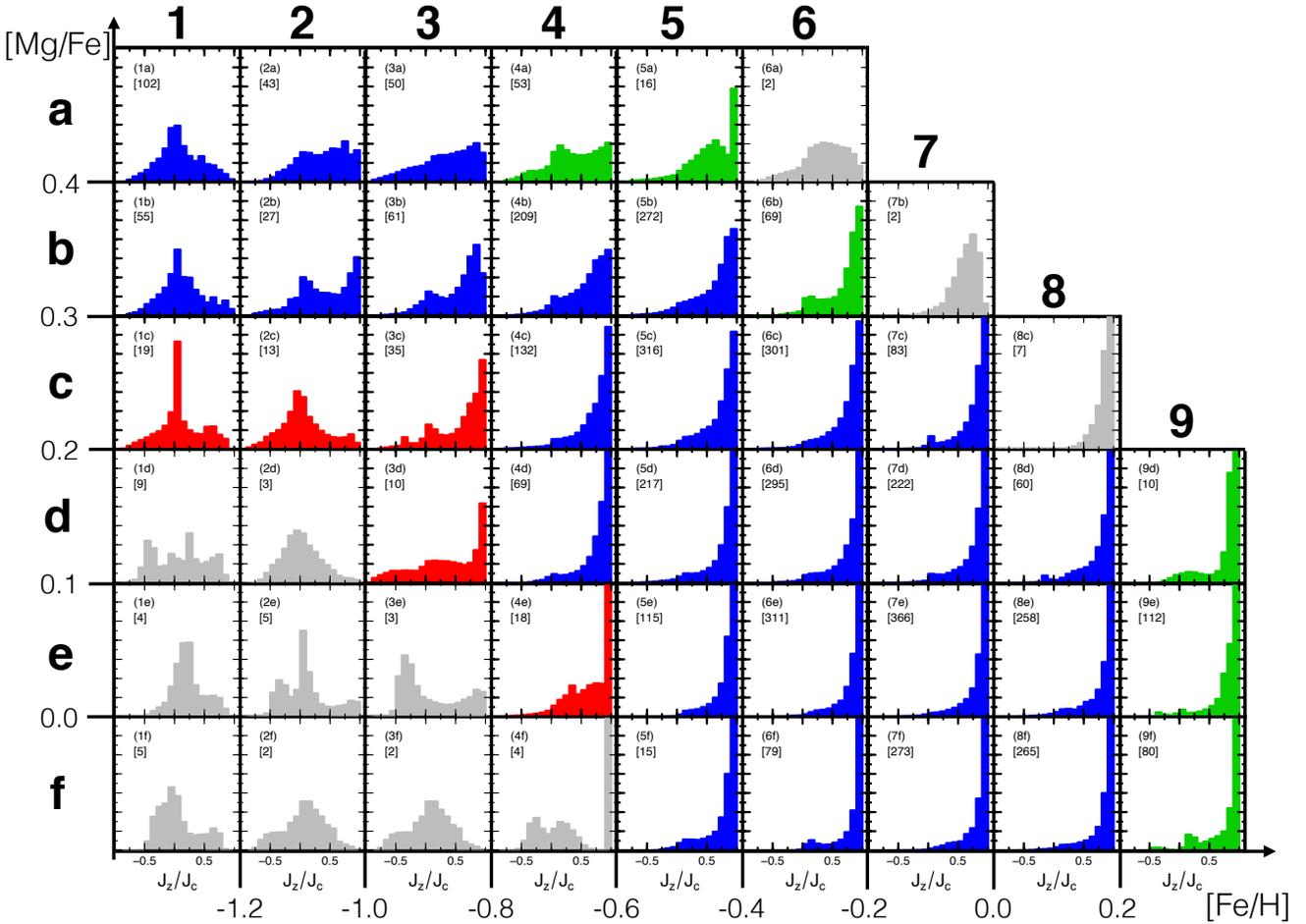}
\caption{The distribution of the vertical component of the specific orbital angular momentum ($\jzjc$) for each $\mgfe-\feh$ bin defined in Figure~\ref{fig-mgfe}. The distributions are computed by summing PDFs for $\jzjc$ for all stars present within a given bin.  This is then normalised such that the total area in a given bin equals one.  The numbers in brackets give the total number of stars in each bin.  The red distributions in 1c, 2c, 3c-d, and 4e show signatures of an accreted component that falls in line with our chemo-dynamical template defined in R14 (namely, $\mgfe<0.3$ and $\jzjc<0.7$).  Additionally, panels 4a, 5a, 6b, 9d-f (shown in green) also contain possible low-$\jzjc$ components.  Panels which contain less than 10 stars, and thus suffer from low-number statistics, are shown in grey.}
\label{fig-jz}
\end{figure*}

\begin{figure*}
\centering
\includegraphics[angle=0,height=0.7\textwidth]{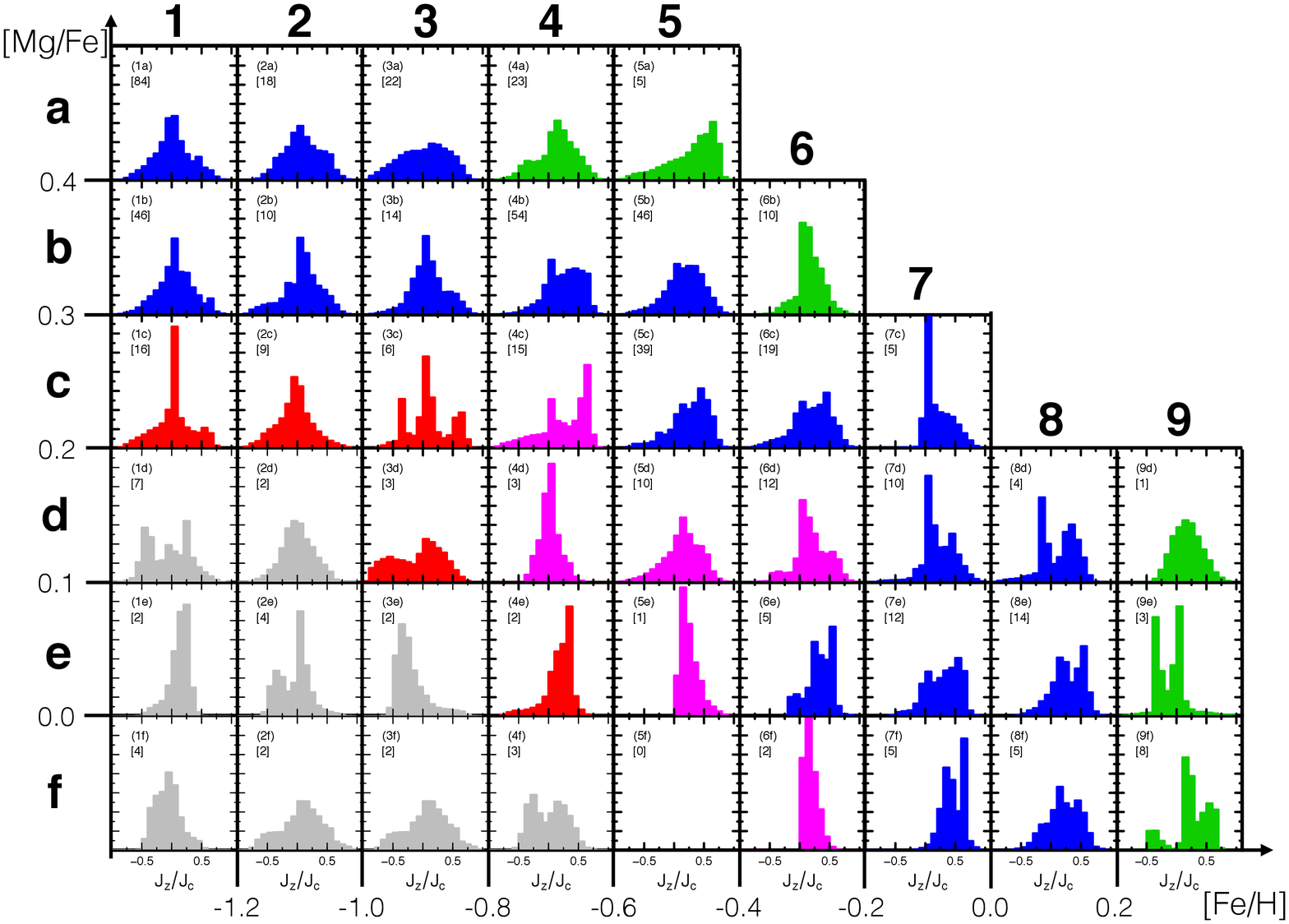}
\caption{The same as Figure~\ref{fig-jz}, but only stars that have probabilities, $P(\jzjc\geq0.7)<0.05$, are plotted.  Panels 4d, 5d-e, 6d, and 6f (now shown as magenta histograms) show a clear low-$\jzjc$ component that was otherwise hidden in Figure~\ref{fig-jz}.  The number in the bracket gives the total number of stars in each bin.}
\label{fig-jzcut}
\end{figure*}

For each $\mgfe-\feh$ bin, we computed the complete distribution of $\jzjc$ by summing the PDFs of individual stars present within that bin.  The summed distribution was normalised such that the total area is equal to one.  The grid of distributions for $\jzjc$ for all abundance bins is shown in Figure~\ref{fig-jz}.  This figure reveals several interesting characteristics:
\begin{enumerate}
\item
The majority of the $\mgfe-\feh$ bins show a primary peak at $\jzjc\sim1.0$ with an extended tail towards low $\jzjc$ values, i.e. the Galactic disc.
\item
The red histograms in panels 1c, 2c, 3c-d, and 4e indicate those bins with $\mgfe<0.3$ (defined in \S\ref{sec-background}) and that have a signature of a distinct low-$\jzjc$ component in addition to the peak at $\jzjc\sim1$.  The low-$\jzjc$ component peaks between 0.00 and 0.25 in all of the bins, except 4e.  Panel 4e shows possible indications of a second component peaking between $\jzjc\sim0.2$ and $0.5$.
\item 
There are also possible signatures of a low-$\jzjc$ component in bins containing stars with high $\mgfe$ ratios at metallicities above $-0.8$~dex (green panels).  This is most evident in panels 4a, 5a, and 6b.  Here, the stars are relatively metal-rich ($-0.8\leq\feh<-0.2$) and $\alpha$-enhanced ($\mgfe\geq0.3$).  Panels 9d-f also contain a handful of low-$\jzjc$ stars with high metallicities, $\feh>0.2$.
\item
Bins that suffer from low-number statistics are shown as grey histograms. Although the statistics for these bins is poor, all of them contain low-$\jzjc$ stars.
\item 
Panels 1a-b, 2a-b, and 3b all show low-$\jzjc$ components. The $\mgfe$ ratios of these bins are consistent with normal high-$\alpha$ halo stars.
\end{enumerate}

The stars identified in panels 1c, 2c, and 3c-d contain the low-$\jzjc$ stars at low metallicity first identified in \S\ref{sec-abund}.  These stars are likely accreted stars, but their low metallicity and low $\jzjc$ suggests they were likely accreted from a low-mass satellite galaxy (see discussion in \S\ref{sec-background}).  In contrast, the more prograde component in panel 4e is in line with the expected specific orbital angular momentum of accreted disc stars that have come from a higher-mass, low-inclination merger. We will discuss the statistical significance of this feature in \S\ref{sec-acc2}.

The panels 4a, 5a, and 6b confirm our detection of a low-$\jzjc$ component at high-$\alpha$ and high metallicity in Figure~\ref{fig-mgsplit}.  These stars have $\mgfe$ that are clearly distinct from those identified in (ii).  The $\mgfe$ ratios of these high-$\alpha$ stars show some similarity to stars observed in the inner disc \cite[e.g.][]{bensby2011} and bulge \citep[e.g.][]{bensby2013,ness2013}, and thus, may have similar origins.  Furthermore, the stars in panels 9d-f could be stars that have migrated outwards from the inner parts of the Galaxy \citep[e.g.][]{kordopatis2015}.  

For the remainder of this paper, we will concentrate on identifying accreted stars and the accreted disc from point (ii).  The low-$\jzjc$ signal for high-$\alpha$, metal-rich stars, described in point (iii), will be discussed in more detail in a forthcoming paper.

Figure~\ref{fig-jz} shows the power of using the distribution of $\jzjc$ in different abundance bins to pinpoint possible substructure in a large sample of stars.  These distributions provide a much clearer picture than using only the abundance trends in Figure~\ref{fig-mgsplit}.  

Still, there could be an accreted component in the disc-dominated bins (blue bins with $\feh\geq-0.8$ in Figure~\ref{fig-jz}), which has been completely hidden by the low-angular momentum tail of the disc component.  In Figure~\ref{fig-jzcut}, we cut away the Galactic disc stars by only plotting those stars which have a probability, $P(\jzjc\geq0.7)<0.05$.  In other words, the stars plotted only have a five percent probability (according to their derived PDF) of having a $\jzjc\geq0.7$.

Figure~\ref{fig-jzcut} reveals stars on low-$\jzjc$ orbits in several disc-dominated bins at metallicities greater than $-0.8$~dex (magenta bins: 4d, 5d-e, 6d, and 6f).  The peak value at $\jzjc\sim0$ is relatively far away from our cut at $\jzjc=0.7$.  Thus, we can be confident that the peak is not just a result of this cut.  These bins, however, contain very few stars and thus it is possible that some of these could be disc stars with very poor measurements.  There are also several other bins (e.g. 7c-d and 8d) that also show indications of low-$\jzjc$ stars.  However, these bins are likely connected to the green histograms identified in Figure~\ref{fig-jz}, which will be discussed in a later paper.

It is clear from Figures~\ref{fig-jz} and \ref{fig-jzcut} that some bins contain very few stars.  This can make it difficult to assess the significance of any detected accreted component. In the next section, we combine bins to reinforce the statistics.

\subsection{Characterizing the accreted population}\label{sec-acc2}

In this section, we explore the evidence for accreted stars present in our sample.  For this analysis, we only considered stars with $\feh<-0.2$.  Above this metallicity, we do not expect to find a significant accreted component (see \S\ref{sec-acc1} as well as Fig.~1 of R14).  In an effort to improve the statistics of the bins associated with accreted stars in Figure~\ref{fig-jz}, we utilize a variant of our original chemical template in R14.  

In R14, we found that the abundance ratios of stars in surviving satellite galaxies reach an upper threshold of $\mgfe\sim0.3$ in the metallicity range of the sample analyzed in that work.  We therefore used a cut at $\mgfe=0.3$ to define high- and low-$\alpha$ stars.  However, a constant cut at $\mgfe=0.3$ across all metallicities is not appropriate for our current Gaia-ESO sample that extends to higher metallicity than R14.  

\begin{figure}
\centering
\includegraphics[height=0.35\textwidth]{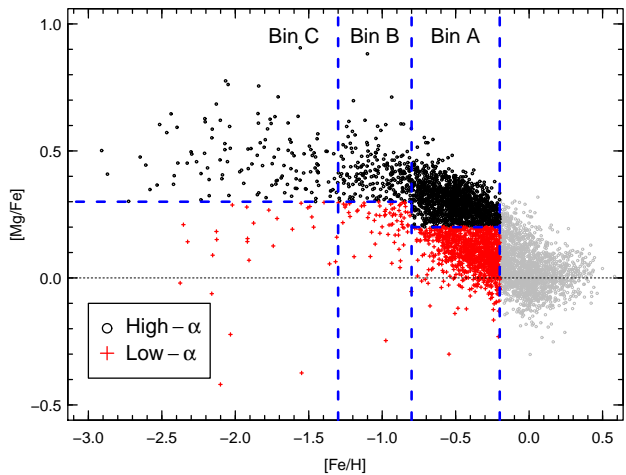}
\caption{The sample plotted in the $\mgfe-\feh$ plane, but now sliced into larger bins, similar to that defined in R14.  Stars are separated into high- and low-$\alpha$ regimes, with a cut at $\mgfe=0.3$ for stars with $\feh<-0.8$.  At larger metallicities, the data were divided by a cut at $\mgfe=0.2$, which was determined using Figure~\ref{fig-jzcut}.  The vertical, dashed lines divide our sample into three metallicity bins, shown as Bin A: $-0.8\leq\feh<-0.2$, Bin B: $-1.3\leq\feh<-0.8$, and Bin C: $\feh<-1.3$.  These bins are also used in Figures~\ref{fig-jzdist}$-$\ref{fig-agedist}.  Stars with $\feh>-0.2$ (grey) are excluded from further analysis.}
\label{fig-bin}
\end{figure}

In Figure~\ref{fig-bin}, we define three $\mgfe-\feh$ bins: Bin A ($-0.8\leq\feh<-0.2$), Bin B ($-1.3\leq\feh<-0.8$), and Bin C ($\feh<-1.3$).  The metallicity range of Bins B and C is similar to that in R14.  We therefore applied the same cut at $\mgfe=0.3$ used in R14 for these bins. For Bin A, this cut is clearly too high according to Figure~\ref{fig-jzcut} where a low-$\jzjc$ component is revealed for most bins with $\mgfe<0.2$. We therefore adopted a cut at $\mgfe=0.2$ for this metallicity bin.

\begin{figure}
\centering
\includegraphics[height=0.6\textwidth]{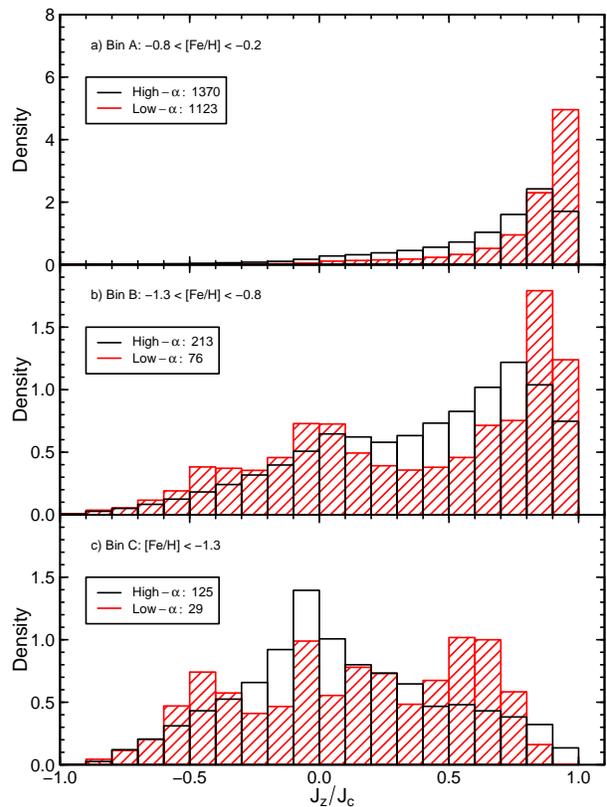}
\caption{Distributions of the vertical component of the specific orbital angular momentum ($\jzjc$) for the three $\feh$ bins defined in Figure~\ref{fig-bin}; (a) Bin A: $-0.8\leq\feh<-0.2$, (b) Bin B: $-1.3\leq\feh<-0.8$, (c) Bin C: $\feh<-1.3$.  The distributions are the sum of the PDF values within a given $\jzjc$ bin, and are normalised such that the total area equals unity.   The total number of stars whose PDF values were summed for each distribution is given in the legend.  The high- and low-$\alpha$ distributions, defined by the cut in [Mg/Fe] shown in Figure~\ref{fig-bin}, are shown as black and red distributions, respectively. }
\label{fig-jzdist}
\end{figure}

We next computed the distribution of $\jzjc$ for the high- and low-$\alpha$ stars (defined by the $\mgfe$ cuts in each bin) in each corresponding $\mgfe-\feh$ bin by summing the $\jzjc$ PDFs of the stars within each given bin.  These are shown in Figure~\ref{fig-jzdist}. Similarly to Figure~\ref{fig-jzcut}, in Figure~\ref{fig-jzdistcut} we explore the possibility that an accreted component could be hidden by the extended tail of the Galactic disc stars by plotting the $\jzjc$ distributions for only stars with $P(\jzjc\geq0.7)<0.05$. This is most useful for Bin A, the high metallicity bin, which has the most substantial disc component. In the following subsections, we explain our findings.

\subsubsection{Bin A: $-0.8\leq\feh<-0.2$}

In this bin, the high- and low-$\alpha$ distributions are very similar (Figure~\ref{fig-jzdist}a) with a peak at $\jzjc\sim1$ and a long tail extending down to $\jzjc\sim0$.  If we cut away the disc distribution as shown in Figure~\ref{fig-jzdistcut}a, we now see a peak around $\jzjc\sim0.1-0.2$, which was hidden in Figure~\ref{fig-jzdist}a. We also see a slight difference between the high-$\alpha$ and low-$\alpha$ components, with the low-$\alpha$ peak slightly shifted to lower $\jzjc$.  A simple two-sample Kolmogorov-Smirnov test yields a p-value$<2\times10^{-16}$, implying that the difference is statistically significant. This suggests that even if the high-$\alpha$ component is just the tail of the disc population (spread to low $\jzjc$ by uncertainties), the low-$\alpha$ component contains something else - a real low $\jzjc$ component from an accreted satellite.

\subsubsection{Bin B: $-1.3\leq\feh<-0.8$}

We find strong evidence for accreted stars in Bin B.  Figure~\ref{fig-jzdist}b shows two distinct peaks in the low-$\alpha$ distribution: one with $\jzjc\sim0.8$ and a low specific angular momentum peak at $\jzjc\sim0$.  This duality is not seen in the high-$\alpha$ distribution.  Figure~\ref{fig-jzdistcut}b further confirms what is seen in Figure~\ref{fig-jzdist}b, that there is a component in the low-$\alpha$ distribution, peaking at $\jzjc\sim0$.

\subsubsection{Bin C: $\feh<-1.3$}

\begin{figure}
\centering
\includegraphics[height=0.6\textwidth]{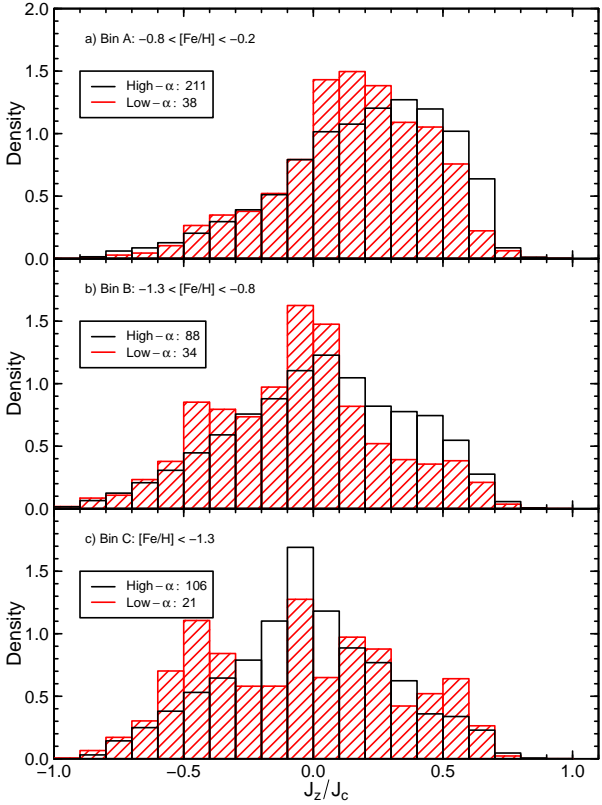}
\caption{The same as Figure~\ref{fig-jzdist}, but only stars with probabilities, $P(\jzjc\geq0.7)<0.05$, are plotted. The three $\feh$ bins are defined in Figure~\ref{fig-bin}; (a) Bin A: $-0.8\leq\feh<-0.2$, (b) Bin B: $-1.3\leq\feh<-0.8$, (c) Bin C: $\feh<-1.3$.  The distributions are the sum of the PDF values within a given $\jzjc$ bin, and are normalised such that the total area equals unity.   The total number of stars whose PDF values were summed for each distribution is given in the legend.  The high- and low-$\alpha$ distributions, defined by the cut in [Mg/Fe] shown in Figure~\ref{fig-bin}, are shown as black and red distributions, respectively. }
\label{fig-jzdistcut}
\end{figure}

At the lowest metallicities, we see accreted stars that have low-$\jzjc$ orbits, as shown in Figure~\ref{fig-jzdist}c. Further, the $\jzjc$ distributions shown in Figure~\ref{fig-jzdistcut}c are almost unchanged, as neither has a disc-like component.  At these metallicities, surviving dwarf galaxies show similar $\mgfe$ ratios to stars in the MW \citep[e.g.][]{tolstoy2009, aden2011,kirby2011}.  We might then expect that the accreted stars should also have similar abundances to the in situ population. We do see, however, subtle differences between the high- and low-$\alpha$ trends. For example, the low-$\alpha$ distribution shows a peak at $\jzjc\sim-0.5$ in both Figure~\ref{fig-jzdist}c and \ref{fig-jzdistcut}c, which is not seen in the high-$\alpha$ distribution.

\subsubsection{The age distributions}

\begin{figure}
\centering
\includegraphics[height=0.6\textwidth]{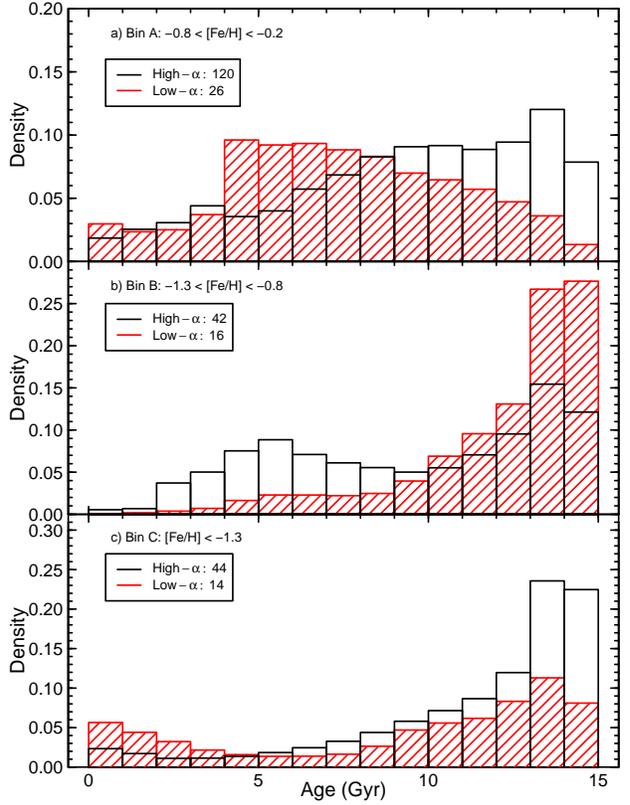}
\caption{Distribution of the ages of stars with $\logg\geq3.5$ and probabilities, $P(\jzjc\geq0.7)<0.05$, in the three $\feh$ bins defined in Figure~\ref{fig-bin}; (a) Bin A: $-0.8\leq\feh<-0.2$, (b) Bin B: $-1.3\leq\feh<-0.8$, (c) Bin C: $\feh<-1.3$.  The total number of stars whose PDF values were summed for each distribution is given in the legend.  The high- and low-$\alpha$ distributions, defined by the cut in [Mg/Fe] shown in Figure~\ref{fig-bin}, are shown as black and red distributions, respectively.}
\label{fig-agedist}
\end{figure}

In Figure~\ref{fig-agedist}, we plot the age distributions of both the high- and low-$\alpha$ stars with $P(\jzjc\geq0.7)<0.05$. As with the $\jzjc$ distributions in Figure \ref{fig-jzdistcut}, these distributions are computed by summing the age PDFs of the stars within a given $\mgfe-\feh$ bin.  There are far fewer stars used to compute the age distributions, since we were limited to only those stars with $\logg>3.5$ (see \S\ref{sec-mage}).  Even though we are strongly affected by low-number statistics, the distributions are still useful as consistency checks of our findings in the $\jzjc$ distributions.

The low-$\alpha$ stars in the low metallicity bins -- Bin B (Figure~\ref{fig-agedist}b) and Bin C (Figure~\ref{fig-agedist}c) -- show predominantly old ages ($>10$~Gyr), consistent with a population of ancient accreted satellites. In contrast, the low-$\alpha$ distribution in the more metal rich Bin A has a peak age at $\sim6-9$~Gyr, albeit with quite a large spread.  Since lower-mass satellite galaxies are expected to grow in metallicity more slowly than the MW, we might expect that metal-rich accreted stars should have younger ages than their more metal-poor counterparts. Thus, it is possible that some of these stars were accreted.  However, such ages are also typical of normal thin and thick disc stars \citep[e.g.][]{haywood2013,bensby2014}. These can extend to low alpha -- especially if the stars are members of the thin disc or have migrated inwards from the outer disc \citep[e.g.][]{Sellwood2002}. It is therefore probable that some of the low-$\alpha$ stars in Figure~\ref{fig-agedist}a are actual disc stars.  This means that even with our probability cuts, this bin still suffers from disc contamination.


\subsubsection{The accreted component}

At metallicities less than $-0.8$~dex (Bins B and C), we have identified a clear signature of an accreted component.  The metallicity range and peak $\jzjc\sim0$ of this component, suggests that the stars were either accreted from one or more low-mass satellite galaxies or one or more LMC-like mass satellites on very high-inclination orbits (see \S\ref{sec-background}, as well as R14). This latter interpretation seems unlikely, however, since such a merger ought to deposit also higher metallicity stars ($\feh>-0.8$) at low $\mgfe$ with low $\jzjc$.  The $\jzjc$ distribution in Figure~\ref{fig-jzdistcut}a (Bin A) shows a higher density for $\jzjc>0.2$ than that shown in Bins B and C.  Further, most of the stars in Bin A have ages similar to Galactic disc stars.  It is therefore likely that the distributions in Bin A are mostly comprised of disc stars and contain few accreted stars.  We conclude that although there may be some accreted stars in Bin A, we do not find a significant accreted component in this metallicity bin, implying that the accreted stars in Bins B and C came from low-mass satellite galaxies.

\section{Mean mass of the merging satellite(s)}\label{sec-merge}

In \S\ref{sec:results}, we confirmed that our sample contains accreted stars.  We suggested that these stars were likely accreted into the halo of the MW from low-mass satellite galaxies.  In this section, we attempt to estimate the mass of the merging satellite galaxies from which the accreted component in our Gaia-ESO sample originated.

In \citet{kirby2013}, the authors compiled the metallicity of individual stars in several dwarf irregular and dwarf spheroidal galaxies in the Local Group.  From this data, they produced a stellar mass$-$metallicity relation for the Local Group dwarf galaxies.  In addition, they combined this with the mass-metallicity relation for more massive galaxies observed in the Sloan Digital Sky Survey \citep{gallazzi2005}.  We adopted this relation to estimate the mass of the merging satellites.  We note, however, that the \citet{kirby2013} mass$-$metallicity relation only depends on the mean metallicity of a dwarf galaxy.  They do not consider the full range of metallicities of the dwarf galaxy, which could provide additional information about the mass distribution and star formation history of the satellite.

In \S\ref{sec-acc2}, we showed that there is an accreted, low-$\alpha$ ($\mgfe<0.3$) component at metallicities $\feh<-0.8$.  In addition, there could be some accreted stars with $\mgfe<0.2$ at metallicities, $-0.8\leq\feh<-0.2$.  If we now consider all of the low-$\alpha$ stars at these metallicities that have $P(\jzjc\geq0.7)<0.05$, the average metallicity is $\left<\feh\right>\sim-1.0$.  Using the mass-metallicity relation in \citet{kirby2013}, this average metallicity corresponds to a satellite galaxy with a stellar mass of ${\rm M}_{*}\sim10^{8.2}$~\msun.  If, on the other hand, we assume that the stars with metallicities $\feh>-0.8$ are all disc stars, then the average metallicity decreases to $\left<\feh\right>\sim-1.3$.  This corresponds to a satellite galaxy with a stellar mass of ${\rm M}_{*}\sim10^{7.4}$~\msun. 

Both of these values for the stellar mass are substantially less than the LMC, which has a stellar mass of ${\rm M}_{*,{\rm LMC}}\sim10^{8.9}$~\msun \citep{vandermarel2002}.  These satellite galaxies would therefore suffer very little disc plane dragging and we would expect them -- when averaging over several mergers -- to deposit their stars on nearly non-rotating orbits. It is reassuring, then, that the $\jzjc$ distribution for these accreted stars in our sample peaks at zero.

It is difficult to quantify exactly how many mergers our accreted star signal corresponds to for two reasons. Firstly, there is a degeneracy between the number and stellar mass of the mergers that is only weakly broken by the metallicity information (see discussion above). Secondly, to properly use information from the Gaia-ESO survey about the number density of stars, we must correct for the survey selection function. This requires a good understanding of the stellar populations both in the Milky Way and the accreted satellites. Such detailed modelling is beyond the scope of this present work.

\section{The quiescent Milky Way}\label{sec-qMW}
 
In \S\ref{sec:results} and \S\ref{sec-merge}, we confirmed our previous results from R14.  A key difference from our previous work, however, is that we do now clearly detect low-mass accretion events. That our analysis finds these, but not any at higher mass proves that the MW has suffered few if any LMC-mass mergers since its disc formed $\sim9$~Gyr ago. As a result, it will not have a significant dark matter disc. This is reinforced by our simulation results in Figure \ref{fig-accsim} where we see that using just the kinematics alone, a massive LMC-like merger would stand out as a distinct peak in $\jzjc$ space that we do not see in the Gaia-ESO data.

Our result that the MW has had no significant mergers since its disc formed is supported by many complementary, albeit more indirect or model dependent, studies of the MW. \cite{hammer2007} investigated the stellar mass and angular momentum of the MW, as well as the $\feh$ of stars in its outer regions, finding that it did not undergo any significant mergers over the last $\sim10$~Gyr.  The kinematic groups identified by \citet{helmi2006} that are likely formed from the accretion of stars from satellites also support this notion, since they have been shown to consist of predominantly old stars \citep[e.g.][]{stonkute2013,zenoviene2014}. Recently, Casagrande et al.~(2015, submitted) used the combination of asteroseismology and spectroscopy to study the vertical age gradient in the Galactic disc.  They found no signatures of over-densities in the age distribution for ages younger than 10~Gyr, which is consistent with a quiescent MW.  In addition, \citet{minchev2014} found that the velocity dispersions of stars in the Galactic disc decrease with increasing $\mgfe$, suggesting that the MW has experienced few late major mergers.  This has recently been confirmed with the Gaia-ESO Survey (Guiglion et al., submitted). Finally, the nearly spherical shape of the dark matter halo, inferred from measurements of the local dark matter density \citep[e.g.][and references therein]{read2014}, and the steep fall-off in the stellar halo density profile \citep[e.g.][]{deason2013} both suggest that the MW had a quiet merger history with an insignificant dark matter disc.

In contrast to the above studies, two over-densities -- the Monoceros Ring and Triangulus Andromeda -- have been identified in the disc. These contain intermediate-age stars \citep[e.g.][and references therein]{carraro2015} and it has been suggested that they could result from accreted satellite galaxies \citep[e.g.][]{2012ApJ...754..101C}. If true, this would imply more recent mergers, at odds with our results here. However, both features could also be the result of disc oscillations \citep[e.g.][]{xu2015} or flaring \citep[e.g.][]{lopez-corredoira2014}. Our results here disfavour a merger origin for these features and lend support to the disc oscillation/flaring scenario.

\section{Conclusions and Future Prospects}\label{sec:conclusions}

In this work, we expanded the chemo-dynamical analysis developed in \citet{ruchti2014} to a large sample of $4,675$ MW stars in the third internal data release from the Gaia-ESO Survey.  Our analysis uncovered a significant component of accreted stars on nearly non-rotating orbits, which were likely accreted from satellite galaxies with stellar masses of $10^{7.4}-10^{8.2}$~\msun.  However, we found no evidence of a significant prograde accreted disc component, which would be the result of one or more $\simgt$LMC-mass mergers.  We conclude that the MW has no significant accreted stellar disc, and thus possesses no significant dark matter disc. This implies that the MW did not experience any substantial ($\simgt 1:10$)  mergers after the formation of the Galactic disc ($\sim9$~Gyr ago), and therefore has had a relatively quiescent merger history.

On the other hand, our template revealed a possible signature of a component of metal-rich ($-0.8\leq\feh<-0.2$), $\alpha$-enhanced ($\mgfe\geq0.3$) stars with low specific vertical angular momentum.  In a follow-up paper, we will investigate the nature of these interesting stars using comparisons to recent state-of-the-art cosmological simulations \citep[e.g.][]{agertz2014}. 

Finally, our results here have interesting implications for the local dark matter density $\rho_{\rm dm}$ and the local dark matter velocity distribution function $f_{\rm dm}$. Firstly, the lack of a dark disc suggests that $f_{\rm dm}$ is near-Gaussian (rather than the double-Gaussian distribution expected in the case of a significant dark disc component; e.g. \citealt{read2014}). Secondly, if $\rho_{\rm dm}$ is found to be substantially higher than spherical extrapolations from the rotation curve would imply, the lack of a rotating dark disc allows us to say with confidence that the dark halo of the MW is locally oblate \citep[e.g.][]{read2014}. Such a locally oblate halo is expected due to contraction of the dark halo in response to disc growth \citep[e.g.][]{1991ApJ...377..365K,2008ApJ...681.1076D,2015arXiv150202916P}, but evidence for it remains elusive \citep[e.g.][]{read2014}.

\section*{Acknowledgements}
GRR, SF, and TB acknowledge support from the project grant ``The New Milky Way" from the Knut and Alice Wallenberg Foundation.  JIR would like to acknowledge support from SNF grant PP00P2\_128540/1.  This work was partially supported by grants ESP2013-41268-R (MINECO) and 2014SGR-1458 (Generalitat of Catalunya).  KL acknowledges support from the European Union FP7-PEOPLE-2012-IEF grant No. 328098.  UH acknowledges support from the Swedish National Space Board (SNSB/Rymdstyrelsen).  LS aknowledges the support of Chile's Ministry of Economy, Development, and Tourism's Millennium Science Initiative through grant IC120009, awarded to The Millennium Institute of Astrophysics, MAS.  This work was partly supported by the European Union FP7 programme through ERC grant number 320360 and by the Leverhulme Trust through grant RPG-2012-541. We acknowledge the support from INAF and Ministero dell' Istruzione, dell' Universit\`a' e della Ricerca (MIUR) in the form of the grant "Premiale VLT 2012". The results presented here benefit from discussions held during the Gaia-ESO workshops and conferences supported by the ESF (European Science Foundation) through the GREAT Research Network Programme.

\bibliographystyle{mn2e}
\bibliography{accref}

\clearpage

\end{document}